\documentclass{aa}  

\usepackage{graphicx}
\usepackage{subcaption}
\usepackage{txfonts}
\usepackage{url}      
\usepackage{bm}        
\usepackage{soul}
\usepackage{supertabular}
\usepackage{hyperref}
\usepackage{xcolor}

\defcitealias{Prole2023}{LP23}
\begin{document}

\title{Halo Mergers Enhance the Growth of Massive Black Hole Seeds}
   \authorrunning{Lewis R. Prole}
   \titlerunning{Halo mergers}

   \author{Lewis R. Prole\inst{1,2}\fnmsep\thanks{E-mail: lewis.prole@mu.ie}
          John A. Regan\inst{1,2},
          Daniel J. Whalen\inst{3},
          Simon C. O. Glover\inst{4}, 
          Ralf~S.~Klessen\inst{4,5,6,7} 
          }
\institute{\inst{1}Department of Physics, Maynooth University, Maynooth, Ireland.\\
   \inst{2}Centre for Astrophysics and Space Science Maynooth, Maynooth University, Maynooth, Ireland.\\
    \inst{3}Institute of Cosmology and Gravitation, Portsmouth University, Dennis Sciama Building, Portsmouth PO1 3FX, UK.\\
    \inst{4}Universit\"{a}t Heidelberg, Zentrum f\"{u}r Astronomie, Institut f\"{u}r Theoretische Astrophysik, Albert-Ueberle-Str.\ 2, 69120 Heidelberg, Germany.\\
    \inst{5}Universit\"{a}t Heidelberg, Interdisziplin\"{a}res Zentrum f\"{u}r Wissenschaftliches Rechnen, Im Neuenheimer Feld 225, 69120 Heidelberg, Germany.\\
    \inst{6}Harvard-Smithsonian Center for Astrophysics, 60 Garden Street, Cambridge, MA 02138, U.S.A..\\
    \inst{7}Elizabeth S. and Richard M. Cashin Fellow at the Radcliffe Institute for Advanced Studies at Harvard University, 10 Garden Street, Cambridge, MA 02138, U.S.A.\\}

   \date{Received ; accepted }

 
  \abstract
   {High redshift observations of 10$^9$ M$_\odot$ supermassive black holes (SMBHs) at $z \sim7$ and `Little Red Dots' that may host overmassive black holes at $z>4$ suggests the existence of so-called heavy seeds (>1000 M$_\odot$) in the early Universe. Recent work has suggested that the rapid assembly of halos may be the key to forming heavy seeds early enough in the Universe to match such observations, as the high accretion rate into the halo suppresses the cooling ability of H$_2$, allowing it to quickly accrete up to the atomic cooling limit of 10$^7$ M$_\odot$ prior to the run-away collapse of baryonic gas within its dark matter (DM) potential, without the need for extreme radiation fields or dark matter streaming velocities.}
   {While the rapid assembly of halos can lead to increased halo masses upon the onset of collapse, it remains unclear if this leads to higher mass BH seeds. As a common route for halos to grow rapidly is via halo-halo mergers, we aim to test what effects such a merger occurring during the initial gas collapse has on the formation of BH seeds.}
   {We perform simulations of BH seed formation in 4 distinct idealised halo collapse scenarios; an isolated 10$^6$ M$_\odot$ minihalo, an isolated 10$^7$ M$_\odot$ atomic halo, the direct collision of two 10$^7$ M$_\odot$ halos and a fly-by collision of two 10$^7$ M$_\odot$ halos. We simulate the collapse of the gas down to scales of $\sim$0.0075 pc before inserting sink particles as BH seeds and capture a further 10 Myr of accretion.}
   {We have shown that halo collisions create a central environment of enhanced density, inside which BH seeds can accrete at enhanced rates. For direct collisions, the gas density peaks are disrupted by the interaction, as the collisionless DM peaks pass through each other while the colliding gas is left in the center, removing the sink particle from its accretion source. When the central density peaks instead experience a fly-by interaction, the sink particle remains embedded in the dense gas and maintains enhanced accretion rates throughout the simulated period when compared to the isolated halo cases. During the total simulated period of 70 Myr, we followed the evolution of the sink particle for the final 10 Myr, with the sink particle spending the final 6 Myr  embedded in the dense, shocked region before the end of the simulation. The final mass of the sink particle achieved a factor of 2 greater in mass than in the isolated atomic halo case, and a factor of 3 greater than the minihalo case, reaching 10$^4$ M$_\odot$ via its 0.03 pc accretion radius. As the maximum halo mass before collapse is determined by the atomic cooling limit of a few times 10$^7$ M$_{\odot}$, the ability of halo-halo mergers to further boost accretion rates onto the central object may play a crucial role in growing SMBH seeds, needed to explain recent observations of seemingly overmassive black holes at high redshift.}


   \maketitle


\section{Introduction} \label{sec:Introduction}

Detections of quasars at redshifts up to $z \sim$ 7.6 with black hole (BH) masses of $\sim 10^{9}$ M$_{\odot}$ (supermassive BHs: SMBHs) (e.g. \citealt{Mortlock2011,Matsuoka2019a,Yang2020,Wang2021,Fan2023,Larson2023}) and now active galactic nuclei (AGNs) at $z \gtrsim$ 8 with masses of $\gtrsim$ 10$^6$ M$_{\odot}$ \citep{Bogdan2024,Bunker2023,Maiolino2024} challenge our understanding of BH formation and growth in the early Universe.

The {\em James Webb Space Telescope} ({\em JWST}) via the JADES, CEERS and UNCOVER surveys have now also revealed the existence of more than 300 compact, red objects at $z$ = 4 - 10 (Little red dots: LRDs) accompanied by broad line emission \citep{Matthee2023,Matthee2024, Harikane2023, Kocevski2023,Feeney2024}, typically indicative of an AGN. 
These objects, due to their unusually strong red colour component in the mid-far infrared may indicate the presence of strong AGN activity. This in turn would indicate that (central) massive black hole (MBH) number densities are much higher than previously thought and hence that MBHs must form much more ubiquitously than previously considered. Unlike typical AGN hosting galaxies, they are, for the most part, not accompanied by X-ray \citep{Ananna2024} or hot dust emission \citep{Williams2024}. Furthermore, when an AGN is assumed present, there is tentative evidence that the SED fits are categorised by high/super Eddington accretion and/or over-massive BH/stellar mass ratios \citep{Volonteri2024,Durodola2024}. All of this (admittedly as of yet somewhat speculative) evidence for higher than expected MBH number densities in the early Universe means that it is imperative that we try to understand how rapid and efficient MBH formation at high-$z$ can develop.



The high-z SMBH population as well as the known high-z AGN dominated galaxies (e.g. GNz-11) and potentially the LRD population seemingly all require early MBH assembly and rapid accretion. As such it is probable that the local environment hosting the embryonic black hole will play a key role in determining the MBH seed mass and, the MBH duty cycle and growth rates. Over the last two decades significant research has taken place to determine the birthplaces for MBH seeds and to determine their formation pathways. While light seeds (<1000 M$_\odot$), born from the remnants of Population III (Pop III) stars are expected to form ubiquitously in metal-free halos (see \citealt{Klessen2023}), it is thought that heavy seeds (>1000 M$_\odot$) must wait for larger halos to emerge. Therefore a strong contender for the formation of heavy seeds are so-called pristine or near-pristine atomically cooled halos \citep{Oh2002a,Bromm2003,Bromm2011,Regan2017,Prole2024a}.

Pristine atomic cooled halos can emerge via a variety of environmental pathways. In almost all cases, ordinary star formation in progenitor halos is suppressed until the halo reaches masses $\gtrsim$ 10$^7$ M$_\odot$ via some suppression mechanism. Examples include immersion in extremely high Lyman-Werner UV fluxes ($J_{21} \gtrsim 1000$ in units of 10$^{-21}$~erg~s$^{-1}$~cm$^{-2}$~Hz$^{-1}$~sr$^{-1}$) that suppress H$_2$ formation (e.g. \citealt{Shang2010,Regan2014,Regan2016,Regan2017,Regan2018,Latif2013,Latif2013b,Latif2014,Suazo2019,Patrick2023}), supersonic baryon streaming motions that prevent gas collapse even if H$_2$ is present \citep{Stacy2011,Greif2011b,Latif2014b,Schauer2017,Hirano2017b}, highly supersonic turbulence due to rare, powerful accretion streams \citep{Yoshida2003, Fernandez2014, Inayoshi2015a, Latif2022a} or some combination of these \citep{Wise2019}. The onset of atomic cooling occurs when a halos virial temperature reaches $\sim$8000 K, with the halo mass required to achieve this being an increasing function with the age of the Universe, going as \citep{Fernandez2014} 
\begin{equation}
\rm M_{halo} = \Big[\frac{T_{vir}}{0.75 \times 1800} \Big(\frac{21}{1 + z}  \Big) \Big]^{3/2} 10^6 M_{\odot}.
\end{equation}
At $z$ = 15, this comes to $\sim$2$\times 10^7$ M$_\odot$. At these masses, catastrophic baryon collapse triggered by the atomic cooling achieves inflow rates of $\sim$ 10$^{-3}$ - 1 M$_\odot$ yr$^{-1}$. At these inflow rates stellar evolution models indicate that Pop III stars with masses of a few 10$^3$ - 10$^5$ M$_\odot$ can emerge and later directly collapse to heavy seed BHs \citep{Hosokawa2013,Woods2017,Haemmerle2018,Hirano2014,Hirano2015,Herrington2023,Nandal2024}.  \\

Heavy seeds are currently favoured as the formation pathway for high redshift SMBHs over normal Pop III stellar BHs (light seeds) with masses of a few tens to hundreds of solar masses \citep{Susa2014,Hirano2014,Hirano2015,Susa2019,Latif2022a} because low-mass Pop III stars are born in low density regions where they can not rapidly grow and can be ejected from their host halos during supernovae (SNe) explosions or gravitational collapse \citep{Whalen2004,Kitayama2004,Whalen2012, Smith2018, Beckmann2019,Pfister2019}.  Although mechanisms such as super- or hyper Eddington accretion have been proposed to overcome these obstacles \citep{Madau2014,Alexander2014,Volonteri2015,Lupi2016,Inayoshi2016}, radiative or mechanical feedback from the BH appears to make sustained periods of such accretion difficult \citep{Johnson2007,Milosavljevic2009,Alvarez2009,Smith2018,Regan2019,Su2023,Massonneau2023}. On the other hand, dynamical processes like runaway stellar mergers in dense nuclear clusters could build up stellar masses to a few 1000 M$_\odot$ (e.g. \citealt{PortegiesZwart2004,Devecchi2009,Katz2015,Reinoso2018,Boekholt2018,Reinoso2023}). Furthermore, feedback in the form of nearby SNe explosions may counter-intuitively aid the growth of light seeds by launching gas towards the accreting objects \citep{Mehta2024}.

The birth of a 10$^4$ - 10$^5$ M$_\odot$ MBH does not by itself guarantee it will reach 10$^7$ - 10$^9$ M$_\odot$ by $z$ = 5 - 10.  The few cosmological simulations that have created a 10$^9$ M$_{\odot}$ BH by $z \sim$ 7 have relied on the "heaviest" seeds in the most extreme environments (e.g. \citealt{Smidt2018}) and find that the MBH must reside at the nexus of rare, unusually powerful accretion streams that grow halos to $\sim$ 10$^{12}$ M$_{\odot}$ in low-shear environments by $z \sim$ 6 \citep{Matteo2012,Feng2014,DiMatteo2017,Tenneti2018,Smidt2018,Zhu2022}.  This fueling scenario can produce a $4 \times 10^7$ M$_{\odot}$ BH by $z$ = 10.4 \citep{Smidt2018}, like that in UHZ1, but only in one or two dozen halos per cGpc$^{-3}$. These are more likely the tip-of-the-iceberg, with the more common outcome being BHs with masses in the range 10$^4$ - 10$^5$ M$_\odot$. It is these BHs with higher number densities and lower seed masses \citep{Regan2024, Mccaffrey2024} that may be the seeds for the LRDs and other AGN hosting galaxies which have number densities in the range $\sim$ 10$^{-5}$ - 10$^{-3}$ cMpc$^{-3}$ \citep{Perez-Gonzalez2024,Greene2024,Kocevski2024}.

As the maximum mass halos can achieve before the onset of collapse is limited by the atomic cooling threshold of a few 10$^7$ M$_\odot$, this also theoretically limits the maximum inflow rates onto the central BH seed. Here we investigate whether rapid halo assembly via atomic halo collisions can further drive up accretion rates and boost the masses of BH seeds compared to isolated halos, ultimately leading to the SMBH population currently being detected at high redshift.

The structure of the paper is as follows: 
in \S \ref{sec:method} we discuss our numerical method, including the simulation code, initial conditions, chemistry solver and our implementation of sink particles. We present the results of our simulations in \S \ref{sec:results} and discuss them in \S \ref{sec:dis}. In \S \ref{sec:caveats}, we highlight a few caveats. We conclude in \S \ref{sec:conclusions}.

\begin{figure*}
\centering
\begin{subfigure}[b]{-4\textwidth}
	 \hbox{\hspace{-6cm} \includegraphics[scale=0.72]{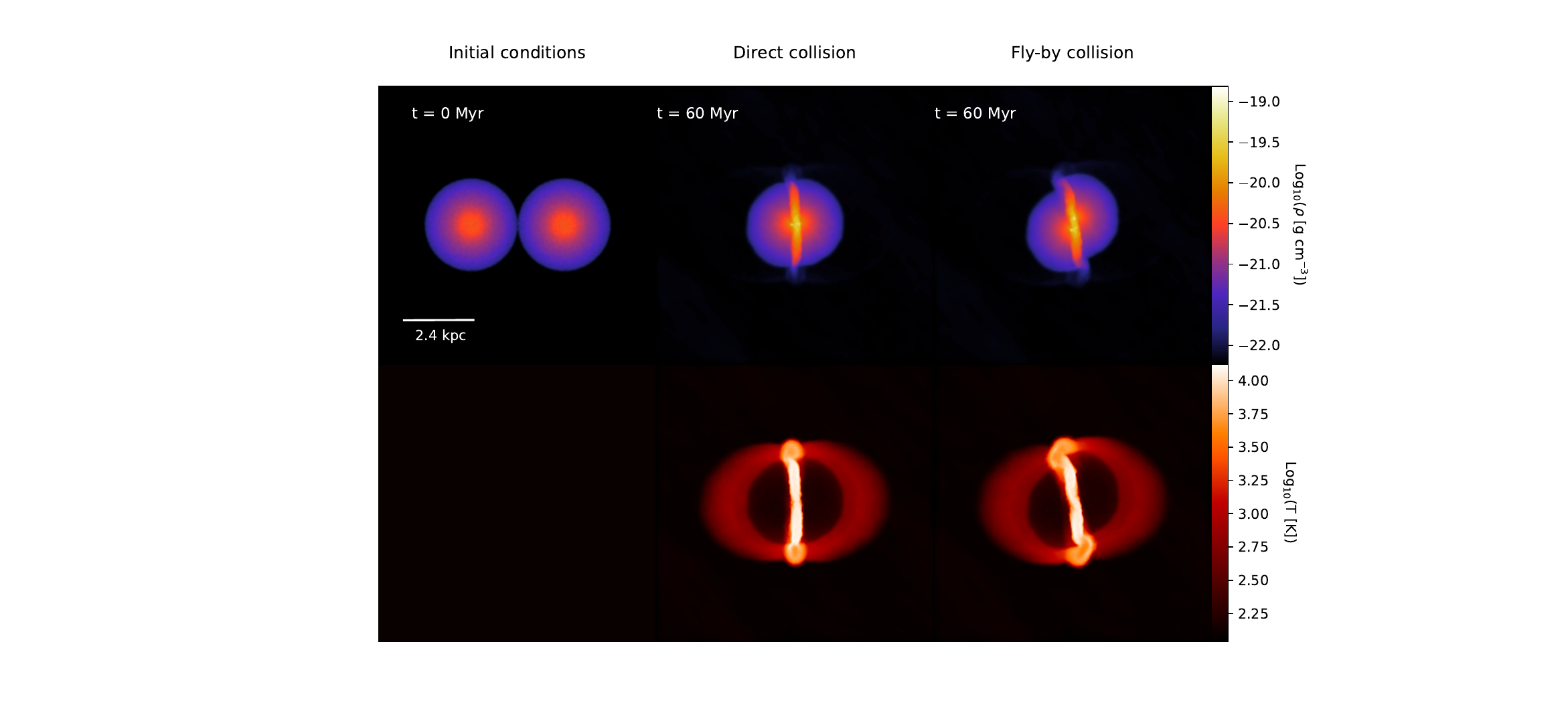}}
\end{subfigure}
    \caption{Density (top) and temperature (bottom) slices of the simulation box (9.6 kpc) for the halo collision scenarios. Left - initial conditions for both direct and fly-by collision simulations. Middle - the direct collision shown just before the formation of the first sink particle. Right - the fly-by collision shown just before the formation of the first sink particle.}
    \label{fig:collision_ics}
\end{figure*}

\begin{figure*}
\centering
\begin{subfigure}[b]{0.55\textwidth}
	 \hbox{\hspace{-1.2cm} \includegraphics[scale=0.85]{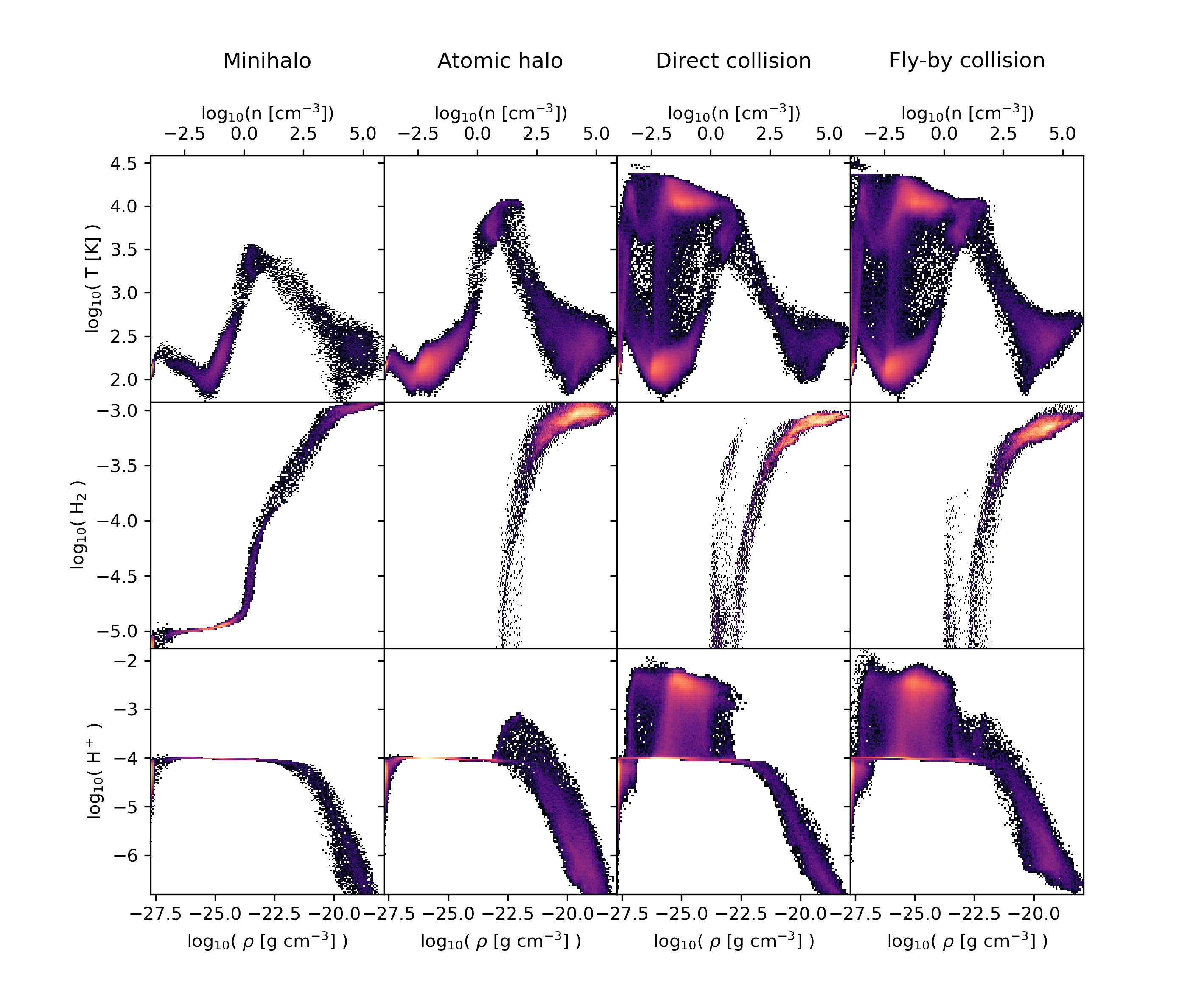}}
\end{subfigure}
    \caption{2D histograms of gas properties across the 4 scenarios at a time just before the formation of the first sink particle, coloured by number of cells. We show temperature (top), H$_2$ abundance (middle) and ionisation fraction (bottom) as a function of density.}
    \label{fig:temp}
\end{figure*}

\begin{figure*}
\centering
\begin{subfigure}[b]{0.55\textwidth}
	 \hbox{\hspace{-0.9cm} \includegraphics[scale=0.8]{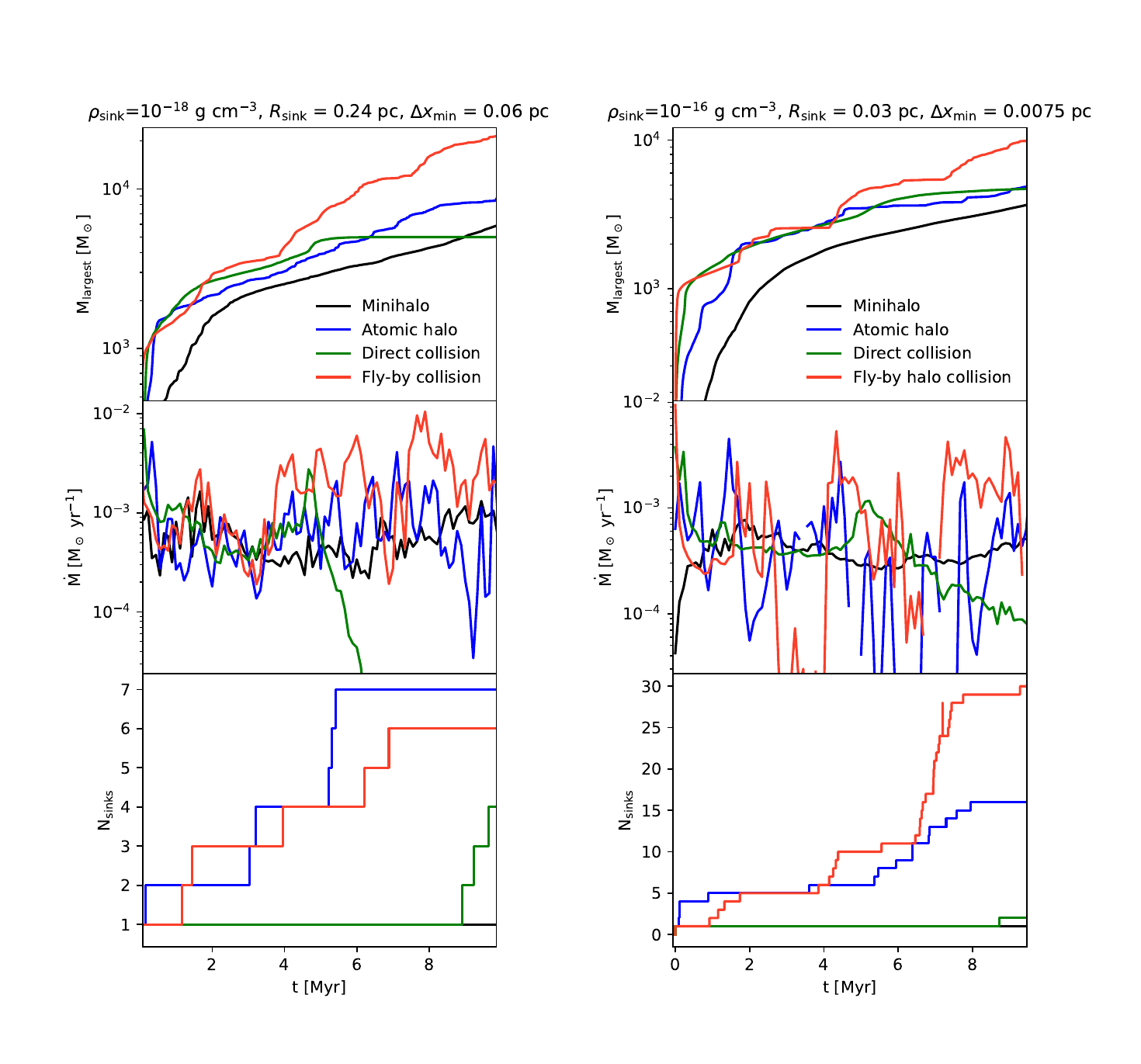}}
\end{subfigure}
    \caption{Sink particle evolution for the 4 halo scenarios as a function of time, shown for both resolutions tested. Top - mass of the most massive sink particle. Middle - accretion rate onto the most massive sink particle, regions with no data indicate periods of zero accretion. Bottom - total number of sink particles formed.}
    \label{fig:growth}
\end{figure*}

\begin{figure*}
\centering
\begin{subfigure}[b]{0.55\textwidth}
	 \hbox{\hspace{-2.5cm} \includegraphics[scale=0.7]{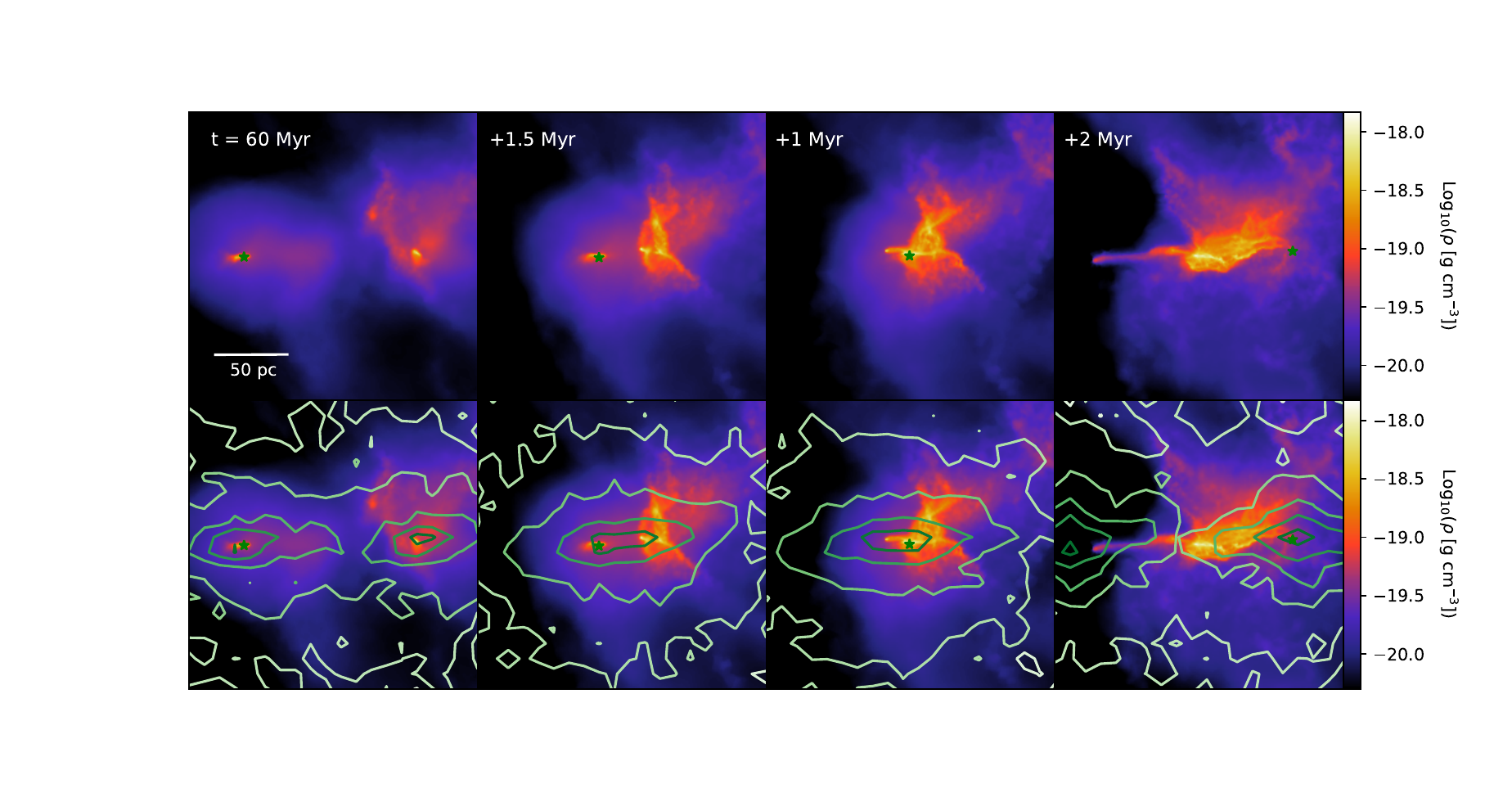}}
\end{subfigure}
    \caption{200 pc zoom slices showing the time evolution of the direct halo collision. Top - density projection of the interaction between the 2 high density peaks of the halos. The green star indicates the sink particle. Bottom - we show the underlying DM distribution by plotting contours of a 2D histogram of the DM particle positions. The slices are shown at times 60 Myr and 1.5, 2.5 and 4.5 Myr later.}
    \label{fig:projection}
\end{figure*}

\begin{figure*}
\centering
\begin{subfigure}[b]{-4\textwidth}
	 \hbox{\hspace{-6cm} \includegraphics[scale=0.72]{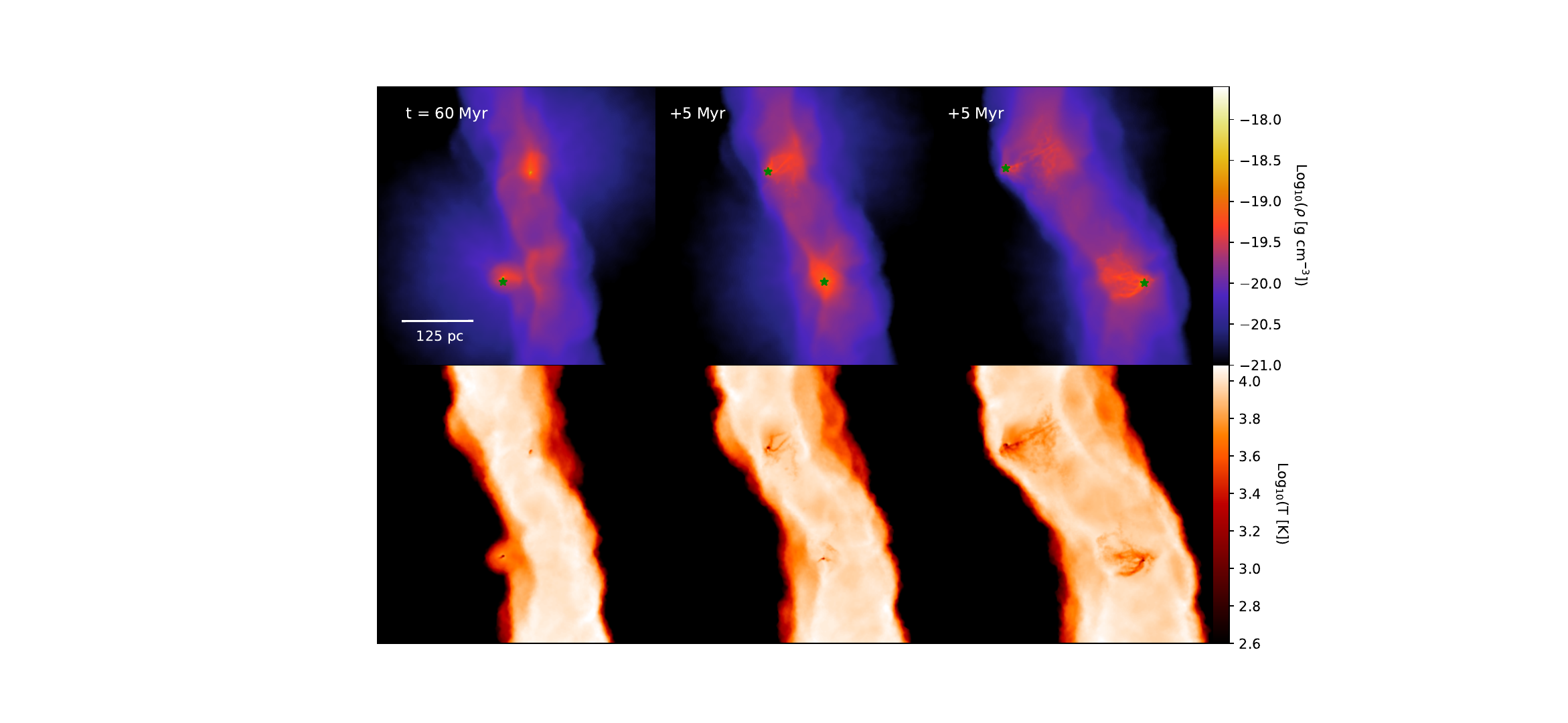}}
\end{subfigure}
    \caption{Density (top) and temperature (bottom) slices of the inner 500 pc of the fly-by collision simulation, shown at the formation of the first sink particle (60Myr), and 5 Myr and 10 Myr later. Sink particles are represented by green star markers.}
    \label{fig:projection2}
\end{figure*}

\begin{figure*}
\centering
\begin{subfigure}[b]{0.55\textwidth}
	 \hbox{\hspace{-0.8cm} \includegraphics[scale=0.8]{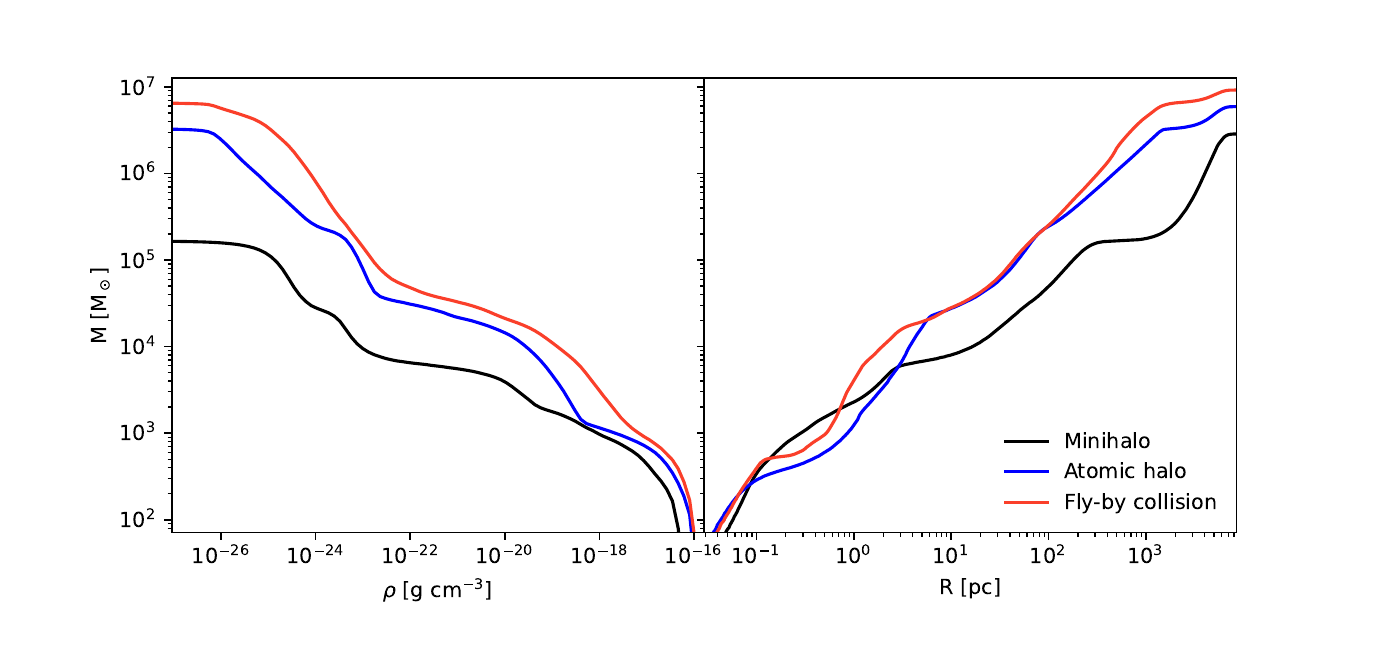}}
\end{subfigure}
    \caption{Mass weighted gas profiles for the isolated halo types and the fly-by collision scenario, compared at 5 Myr after the formation of the first sink particle i.e. when the sink particle is deeply embedded in the dense, shocked region (see middle panel of Figure \ref{fig:projection2}). LHS - mass of gas at or above density thresholds and a function of the density threshold. Right - cumulative mass as a function of radius from the sink particle.}
    \label{fig:radial}
\end{figure*}

\begin{figure}
\centering
\begin{subfigure}[b]{0.55\textwidth}
	 \hbox{\hspace{-0.8cm} \includegraphics[scale=0.8]{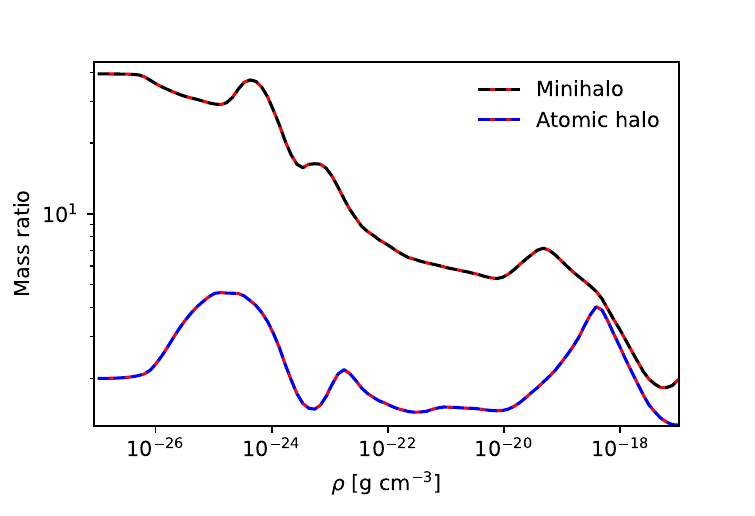}}
\end{subfigure}
    \caption{Total gas mass existing above threshold densities, expressed as a ratio of the gas mass produced in the fly-by collision scenario, shown at 5 Myr after the creation of the first sink particle.}
    \label{fig:ratio}
\end{figure}

\section{Numerical method}
\label{sec:method}
\subsection{{\sc Arepo}}
The simulations presented here were performed with the moving mesh code {\sc Arepo} \citep{Springel2010} with a primordial chemistry set-up described in \S \ref{sec:chem}. {\sc Arepo} combines the advantages of adaptive mesh refinement (AMR: \citealt{Berger1989}) and smoothed particle hydrodynamics (SPH: \citealt{Monaghan1992}) with a mesh made up of a moving, unstructured, Voronoi tessellation of discrete points. {\sc Arepo} solves hyperbolic conservation laws of ideal hydrodynamics with a finite volume approach, based on a second-order unsplit Godunov scheme with an exact Riemann solver. Automatic and continuous refinement overcome the challenge of structure growth associated with AMR (e.g. \citealt{Heitmann2008}). 

\subsection{Initial conditions}
\label{sec:ics}
We model our primordial halos as Navarro-Frenk-White (NFW) \citep{Navarro1996,Navarro1997} dark matter distributions, normalised to give the desired mass within the desired radius. We model the dark matter as particles of mass 100 M$_\odot$ with a gravitational softening length of 0.1 pc. The baryonic gas component is initialised as an isothermal sphere ($\rho \propto$ r$^{-2}$) at 100 K with a central density of 10$^{-25}$ g cm$^{-3}$ spanning out to its Jeans length of $\sim$360 pc. The central density was chosen to represent a halo on the brink of collapse (e.g. \citealt{Greif2008,Schauer2021,Prole2023}). The halos are placed in a background medium of density 10$^{-28}$ g cm$^{-3}$. We take the appropriate chemical abundances of H$_2$, H$^{+}$, D$^{+}$ and HD from \cite{Greif2008} as $x_{\text{H}_{2}}=10^{-5}$, $x_{\text{H}^{+}}=10^{-4}$, $x_{\text{D}^{+}}=2.6\times$10$^{-9}$ and $x_{\text{HD}}=3\times$10$^{-8}$, respectively. The simulation box has side length 9600 pc and is seeded with a randomly generated \cite{Burgers1948} turbulent velocity field. The turbulence at these spatial scales and initial gas densities has been shown through cosmological simulations to be supersonic (e.g. \citealt{Prole2023}), we therefore scale the velocity field to to give an rms velocity of 2 times the sound speed as $v_{\rm rms}$ = 1.82 km s$^{-1}$. We set up 4 controlled numerical tests at $z$ = 15 as follows:
\begin{enumerate}

    \item At our adopted redshift of $z$ = 15, minihalos typically collapse once they reach a mass of $\sim$10$^6$ M$_\odot$ in the absence of a LW radiation field. We therefore set our control set-up as the isolated collapse of a minihalo with 10$^6$ M$_\odot$ in DM and use the cosmological baryon to DM mass ratio of 0.15 to give a gas component of 1.5$\times$10$^5$ M$_\odot$. Generating this gas mass with a 100 K isothermal density profile of central density 10$^{-25}$ g cm$^{-3}$ requires a halo radius $R_{\rm mini}$ = 300 pc.

    \item At our adopted redshift of $z$ = 15 and in the presence of a strong LW radiation field, collapse will be delayed until the halo reaches a few times 10$^7$ M$_\odot$ where atomic cooling can facilitate the collapse (e.g. \citealt{OShea2008}). We therefore create an isolated atomic halo with 2$\times$10$^7$ M$_\odot$ in DM with a baryonic component of 3$\times$10$^6$ M$_\odot$, requiring a radius $R_{\rm atomic}$ = 1600 pc. We set the LW background to a value of $J_{21}$ = 10 in units of 10$^{-21}$~erg~s$^{-1}$~cm$^{-2}$~Hz$^{-1}$~sr$^{-1}$ to allow the gas to reach the atomic cooling limit of 8000 K before collapsing. 

    \item We simulate the merger of two of the above described atomic halos with a halo-halo collision in the presence of a $J_{21}$ = 10 LW field. The halo centres were given positional offsets from the box center of $\pm R_{\rm atomic}$ along the $x$ axis. As the simulation timestep becomes shorter as the density of the gas increases, the period of time we can simulate once the gas reaches the maximum threshold density is significantly shorter than the simulated period prior to this point. It is therefore computationally convenient for the gas to reach its maximum threshold density near to the time when the centres of the halos meet. We experimentally determine the these halos will collapse in roughly 60 Myr. We therefore set the initial collision velocity of the each halo in the $x$ direction to be $\pm$25 km s$^{-1}$. Such velocities are consistent with those seen in previous studies of colliding flows (e.g. \citealt{Latif2021}).

    \item A more likely scenario than a head on collision is an indirect halo collision resulting in a fly-by event between the central density peaks of the halos. We therefore also simulate a collision with a small angular offset in the collisional velocity vectors. The  left and right halos are given angular velocity offsets from the $x$ axis of $+\pi$/24 and -$\pi$/24, respectively. This gives an impact parameter (point of closest approach) of $\sim$200 pc.
    
\end{enumerate}
\subsection{Chemistry}
\label{sec:chem}
Collapse of primordial gas is closely linked to the chemistry involved (e.g. \citealt{Glover2006,Yoshida2007, Glover2008, Turk2011,Klessen2023,Prole2024}). We therefore use a fully time-dependent chemical network to model the gas. We use the treatment of primordial chemistry and cooling originally described in \cite{Clark2011}, but with updated values for some of the rate coefficients, as summarised in \cite{Schauer2019}. The network has 45 chemical reactions to model primordial gas made up of 12 species: H, H$^{+}$, H$^{-}$, H$^{+}_{2}$ , H$_{2}$, He, He$^{+}$, He$^{++}$, D, D$^{+}$, HD and free electrons. Optically thin H$_{2}$ cooling is modelled as described in \citet{Glover2008}: we first calculate the rates in the low density ($n \rightarrow 0$) and LTE limits, and then smoothly interpolate between them as a function of $n / n_{\rm cr}$, where $n_{\rm cr}$ is the H$_{2}$ critical number density above which collisions are so frequent that they keep the populations close to their LTE values. To compute the H$_{2}$ cooling rate in the low density limit, we account for the collisions with H, H$_{2}$, He, H$^{+}$ and electrons. To calculate the H$_{2}$ cooling rate in the optically thick limit, we use an approach based on the Sobolev approximation \citep{Yoshida2006, Clark2011}. Prior to the simulation, we compute a grid of optically thick H$_{2}$ cooling rates as a function of the gas temperature and H$_{2}$ column density. During the simulation, if the gas is dense enough for the H$_{2}$ cooling to potentially be in the optically thick regime ($\rho > 2 \times 10^{-16} \: {\rm g \: cm^{-3}}$), we interpolate the H$_2$ cooling rate from this table, using the local gas temperature and an estimate of the effective H$_{2}$ column density computed using the Sobolev approximation. In addition to H$_{2}$ cooling, we also account for several other heating and cooling processes: cooling from atomic hydrogen and helium, collisionally-induced H$_{2}$ emission, HD cooling, ionisation and recombination, heating and cooling from changes in the chemical make-up of the gas and from shocks, compression and expansion of the gas, three-body H$_{2}$ formation and heating from accretion luminosity. For reasons of computational efficiency, the network switches off tracking of deuterium chemistry\footnote{Note that HD cooling continues to be included in the model.} at densities above 10$^{-16}$~g~cm$^{-3}$ and instead we assume that the ratio of HD to H$_{2}$ at these densities is given by the cosmological D to H ratio of 2.6$ \times $10$^{-5}$. The adiabatic index of the gas is computed as a function of chemical composition and temperature with the {\sc Arepo} HLLD Riemann solver.

\subsection{Sink particles}
\label{sec:sinks}
The simulation mesh must be refined during a gravitational collapse to ensure the local Jeans length is resolved. During the collapse, we resolve the mesh such that the Jeans length \citep{Jeans1902,Bonnor1957} is resolved by 16 cells up to a threshold density $\rho_{\rm sink}$, above which a sink particle \citep{Bate1995} is introduced to represent the dense gas, preventing artificial instability in cells whose Jeans lengths continue to decrease below the minimum cell length. 

We have shown in previous studies that the degree of fragmentation in metal-free gas is highly dependent on numerical resolution, with higher resolution simulations exhibiting more fragmentation \citep{Prole2022,Prole2022a}. We therefore expect that the masses achieved by the sink particles in this study would be lower if we employed higher resolutions. What is not known is whether any trends discovered during the halo collisions in the current experiment are resolution dependent. To that end, we repeat the simulations with threshold densities of  10$^{-18}$ g cm$^{-3}$ and 10$^{-16}$ g cm$^{-3}$. We choose our sink particle accretion radius to be the Jeans length at the sink creation density, calculated using the temperature-density relation provided in \cite{Prole2022}. We set the minimum gravitational softening length for gas and sink particles as $L_{\rm soft}$ = $R_{\rm sink}/2$ and the minimum cell length $L_{\rm min}$ = $R_{\rm sink}/4$. The sink particle parameters are given in Table \ref{table:sinks}. At these resolutions, our sink particles likely represent a dense stellar cluster rather than a single massive star (e.g. \citealt{Prole2023}). The higher resolution simulation ($\rho_{\rm sink}$ = 10$^{-16}$ g cm$^{-3}$) is henceforth taken to be the fiducial case, providing the data in all of the figures unless stated otherwise.

\begin{table}
	\centering
	\caption{Simulation parameters. From left to right, we provide the sink particle creation density, accretion radius, minimum gravitational softening length and minimum cell volume.}
	\label{table:sinks}
	\begin{tabular}{cccc} 
		\hline
		 $\rho_{\rm sink}$ [g cm$^{-3}$]& R$_{\rm sink}$ [pc] & L$_{\rm soft}$ [pc] & $\Delta x_{\rm min}$ [pc] \\
		\hline
		 10$^{-18}$ & 0.24 & 0.12 & 0.06\\
		 10$^{-16}$ & 0.03  & 0.015 &  0.0075   \\
		\hline
	\end{tabular}
\end{table}

Our sink particle implementation was introduced in \cite{Wollenberg2020} and \citet{Tress2020}. A cell is converted into a sink particle if it satisfies three criteria:

   \begin{enumerate}
      \item The cell reaches a threshold density.
      \item It is sufficiently far away from pre-existing sink particles so that their accretion radii do not overlap.
      \item The gas occupying the region inside the sink is gravitationally bound and collapsing.
   \end{enumerate}
   Likewise, for the sink particle to accrete mass from surrounding cells, the cell must meet two criteria: 
      \begin{enumerate}
      \item The cell lies within the accretion radius.
      \item It is gravitationally bound to the sink particle.
   \end{enumerate}
   A sink particle can accrete up to 90$\%$ of a cell's mass, above which the cell is removed and the total cell mass is transferred to the sink. 
The sink particle treatment also includes the accretion luminosity feedback from \cite{Smith2011} and the sink particle merger routine from \cite{Prole2022}.

\section{Results} 
\label{sec:results}
In Figure \ref{fig:collision_ics} we show density and temperature slices of the initial conditions for the direct and fly-by collision scenarios, along with respective slices of both collisions at a point just before the formation of the first sink particle. While the initial conditions are idealised isothermal spheres, as the outskirts of the halos collide in the $x$ direction, the overlapping regions create a plane of enhanced density in the $y-z$ plane, which is shock heated up to the atomic cooling limit of $\sim$10$^4$ K. 

In Figure \ref{fig:temp} we compare the density-temperature profile of the merging halos with that of the isolated mini and atomic halos at a point just before the formation of the first sink particle. This occurred roughly 60 Myr after the simulation started in both the collision scenarios and the isolated atomic halo scenario, while it occurred after around 45 Myr in the isolated minihalo case. Referring to the top row of panels, gas falling into the minihalo follows the well established rise to 10$^3$ K and fall to 10$^2$ K indicative of efficient H$_2$ production and cooling. Likewise, the atomic halo behaves as in previous studies; the H$_2$ abundance is reduced by the external LW field, allowing it to grow to 10$^7$ M$_\odot$ where temperatures reach the atomic cooling limit of 10$^4$ K and Lyman-$\alpha$ cooling facilitates the proceeding collapse. For the halo merger scenarios, low density gas is also heated up to the atomic limit by the highly supersonic motions of the gas and the shocks created by the collision. There are also elevated H$_2$ fraction features at densities around 10$^{-24}$ g cm$^{-3}$ (middle row), likely caused by the increase in fractional ionisation (bottom row) produced by the shocks during the collision, as the H$_2$ formation rate is dependent on the availability of free electrons. 

In Figure \ref{fig:growth} we show the growth of the most massive sink particle in all 4 simulations, accross both resolutions tested. By the end of the simulations, the direct collision scenario did not produce a more massive sink particle than isolated atomic halo case, regardless of resolution. In both cases, the accretion rate onto the sink experiences a brief (1-2 Myr) spike followed by sharp decline, occurring around 5 Myr after its creation. To explore why this happens, we show gas density slices of the interaction between the high density centers of the colliding halos, along with the underlying DM distributions in Figure \ref{fig:projection}. The DM consists of collisionless particles, hence the two central dark matter halos pass through each other during the interaction. On the other hand, the high density gas from the two halos collides and is ripped from its respective DM peak. The bulk of the gas is left in the center, pulled equally in both directions by the diverging DM distributions, while the collisionless sink particle is pulled through and subsequently out of the high density gas with the right-hand-side (RHS) DM potential, causing the accretion peak and subsequent drop-off, as the sink particle is removed from its accretion source. This process is mirrored on the left-hand-side (LHS) of the interaction, with the high density gas leaving with the LHS DM peak despite having not formed a sink particle yet,  showing that the ejection is not a numerical artifact caused by the sink particle routine. The interaction is comparable with observations of the `Bullet Cluster' 1E0657-56 \citep{Markevitch2002,Markevitch2007} which has been used as evidence for the existence of DM due to the offset between the X-ray emission and the nearby gravitational lensing centers \citep{Clowe2006}. Numerical works have recreated the structure of the Bullet Cluster by simulating a encounter between a subcluster and its
main cluster (e.g. \citealt{Mastropietro2008}). Here the collision-dominated hot plasma was separated from the collisionless stellar and DM components, just as we see in our direct collision simulation.

While the direct collision hindered accretion, the remaining 3 scenarios all produced sustained accretion throughout the simulations, with the fly-by collision providing the best conditions for growth. It has been shown in previous works that larger halo masses increase the accretion onto the central objects \citep{OShea2008, Bromm2003, Bromm2011, Wise2019, Latif2022, Prole2024}. In our high resolution (low resolution) case, we see a factor of 1.3 (1.5) boost in mass when comparing the isolated minihalo with the more massive atomic halo case, growing to 3500 and 4500 (5700 and 8300) M$_\odot$ by the end of the simulation, respectively. On top of this, we also see a further enhancement of a factor of 2.2 (2.5) during the fly-by collision, giving a factor of 2.9 (3.6) times the mass achieved by the minihalo, giving a final mass of 10$^4$ (2.08$\times$10$^4$) M$_\odot$ by the end of the simulation. We investigate the reason for this by showing density slices of the fly-by encounter in Figure \ref{fig:projection2}. The accretion onto the sink particle was initially similar to that of the isolated atomic halo, as the sink particle formed outside of the high density, shocked region. However, once the sink particle entered the hot (atomic cooling) dense gas after $\sim$4 Myr, it was able to accrete more efficiently and the growth gains compared to the isolated halo scenario became more apparent. The sink particle remained in the shocked region throughout the remainder (6 Myr) of the simulation. The result that objects accreting within the dense, shocked region of the collision is seemingly resolution independent.

In Figure \ref{fig:radial} we compare the distribution of mass as functions of density and distance from the sink particle at t = 5 Myr after the formation of the first sink particle i.e. when the sink particle is deeply embedded in the dense, shocked region (see middle panel of Figure \ref{fig:projection2}). From the left panel we see that the collision produces increased masses across the entire simulated range of densities. The right panel shows the cumulative mass as a function of radius from the most massive sink particle, showing that the mass gains reside from $\sim$100 pc outward, but also from within the inner few pc. We investigate this further by expressing the enclosed mass-density profiles of the isolated halos as a ratio of the masses produced in the fly-by collision in Figure \ref{fig:ratio}. From this, we see the mass gains compared to the isolated atomic halo mainly reside in the low density range of 10$^{-26}$-10$^{-24}$ g cm$^{-3}$ and the high density range of 10$^{-19}$-10$^{-17}$ g cm$^{-3}$. The increased mass at these densities is more than could be explained by the doubling of gas given from simply having 2 halos present during the merger instead of a single halo. From Figure \ref{fig:temp}, it is clear that the increase in lower density mass is made up of the hot shocked gas in the collisional plane. 

While we did not follow the simulation until the sink particle emerged from the LHS of shocked region, the time periods simulated here are already too long to neglect supernovae (SNe) feedback. For reference, during the 6 Myr period the sink particle spends in the shocked region, any unresolved fragmentation and resulting star formation would result in SNe feedback from any stars with masses higher than $\sim$20 M$_\odot$, while even the highest resolution simulations (hence the least conductive to massive star formation due to fragmentation-induced starvation) show the formation of Pop III stars with masses of a few tens of M$_\odot$ \citep{Prole2023,Prole2024a}. A recent study by \cite{Mehta2024} showed that SNe feedback can actually aid accretion in some cases by shocking the gas towards other accreting bodies. It is therefore not obvious if further shocking the dense collisional plane via the inclusion of SNe feedback would help or hinder BH seed growth. 

The realised boost to BH seed accretion during the collision is important because enhanced accretion is typically only attainable by increasing the mass of the halo when considering isolated halo collapses. This has the limitation that the maximum halo mass that can be achieved before collapse is set by the onset of atomic cooling, meaning halos can not resist collapse past masses of a few $\sim$10$^7$ M$_\odot$, even in the presence of extremely high LW radiation intensities \citep{Prole2024a}, steaming velocities \citep{Schauer2021} or colliding flows \citep{Latif2022}. Boosting the initial seed masses via halo mergers may therefore be the only mechanism that can overcome this physical barrier to BH seed growth.


\section{Discussion}
\label{sec:dis}
In a $\Lambda$CDM context, structure formation occurs via bottom-up hierarchical growth with frequent mergers (e.g. \citealt{Peebles2015}). The first halos are expected to start collapsing from $z$$\sim$30 onwards \citep{Klessen2023} and we now have observations of 10$^9$ M$_\odot$ SMBHs existing by $z$$\sim$11 \citep{Bunker2023,Maiolino2024}. While isolated halos can have their accretion supply cut off when stellar feedback evacuates gas from the halo, mergers can continue to bring fresh gas into the halo. Through multiple halo mergers similar to that of this work, the central black hole seed could continue to grow at accelerated rates for significantly extended periods of time when compared to an isolated collapse, giving them a better chance at achieving observed SMBH masses. 

At $z$$\sim$11, the age of the Universe is roughly 400 Myr. Cosmological simulations show that atomic cooling halos collapse by $z$$\sim$15 \citep{Xu2016,Wise2019,Regan2020a,Latif2022}, giving them on the order of 100 Myr to undergo multiple mergers. While the simulated period in this investigation took roughly 70 Myr, the initial conditions represented the very early stages of collapse and the velocities were fine tuned to match the collapse time. In reality, mergers can occur at any stage during or after the initial collapse and multiple mergers can take place simultaneously. Additionally, it is unlikely that merging halos would be at similar stages of their collapse, and massive halos would likely merge with multiple smaller halos which are yet to have had their collapse triggered by reaching the mass threshold for collapse. The result is that halo-halo mergers can occur over a wide range of time scales which allows for many mergers during the 100 Myr time window.

This experiment is the motivation for future work where we will investigate BH seed formation within halos undergoing rapid mergers in a more realistic cosmological setting. Halos experiencing rapid mergers are likely to be compact and perhaps host unusual star formation histories which could shed light on the formation of LRDs and GNz-11 type galaxies.

\section{Caveats} 
\label{sec:caveats}
The initial conditions presented here are idealised to test the specific concept that halo mergers can facilitate high mass BH seed formation. As a result, the halos lack significant substructure. In future work, we aim to test the results of this work with initial conditions more in line with $\Lambda$CDM cosmology, by sampling the expected primordial power spectrum and performing cosmological simulations.

We have not included magnetic fields in these simulations. While studies of primordial magnetic fields suggest that they can increase the mass of protostars (e.g. \citealt{Peters2014,Turk2012,Saad2022,Hirano2022,Stacy2022}) and SMBHs \citep{Latif2013a,Latif2014a}, the fields have no effect when they are properly resolved, distributing the magnetic energy from the small-scale turbulent dynamo \citep{Schober2012} to smaller spatial scales \citep{Prole2022a}. 

Ideally, we would resolve up to the protostellar formation density of 10$^{-4}$ g cm$^{-3}$ \citep{Greif2012}, which is currently computationally unfeasible for the period of time we aimed to simulate. Failure to resolve protostellar densities means we have not achieved numerical convergence. However, the important factor in these simulations is not the specific sink particle masses, but rather the relative difference between the masses produced in an isolated collapse versus the merger scenario.

\section{Conclusions}
\label{sec:conclusions}
This investigation aimed to test to what extent halo-halo mergers affect the growth of SMBH seeds. To that end, we have performed simulations of the collapse of an isolated minihalo and atomic halo, and compared them to direct and fly-by halo-halo collisions. During the total simulated period of 70 Myr, we followed the evolution of the sink particle for the final 10 Myr, with the sink particles in the collision set-ups spending the final 6 Myr  embedded in the dense, shocked region before the end of the simulations. We have shown that:
\begin{enumerate}
    \item Halo collisions create a central environment of enhanced density, inside which BH seeds can accrete at enhanced rates.
    \item For direct collisions, the gas density peaks are disrupted by the interaction, as the collisionless DM peaks pass through each other while the colliding gas is left in the center, removing the sink particle from its accretion source.
    \item   When the central density peaks instead experience a fly-by interaction, the sink particle remained embedded in the dense gas and maintained enhanced accretion rates when compared to the isolated halo cases.
    \item  For the fly-by simulation, the final mass of the sink particle achieved a factor of 2 higher in mass than the isolated atomic halo, and a factor of 3 higher than the minihalo case, reaching 10$^4$ M$_\odot$ via the $\sim$0.03 pc accretion radius of the sink particle.
    \item  As the maximum halo mass before collapse is determined by the atomic cooling limit of a few times 10$^7$ M$_{\odot}$, the ability of halo-halo mergers to further boost accretion rates onto the central object may play a crucial role in growing SMBH seeds, needed to explain recent observations of seemingly over-massive black holes at $z$$\sim$7. 
\end{enumerate}
While this work utilised idealised initial conditions, it serves as the motivation for future investigations where we will study halo collisions in a cosmological context, using realistic initial conditions more in line with $\Lambda$CDM cosmology.

\begin{acknowledgements}
LP and JR acknowledge support from the Irish Research Council Laureate programme under grant number IRCLA/2022/1165. JR also acknowledges support from the Royal Society and Science Foundation Ireland under grant number URF\textbackslash R1\textbackslash 191132. 
\ \
The simulations were performed on the Luxembourg national supercomputer MeluXina.
The authors gratefully acknowledge the LuxProvide teams for their expert support.
\ \ 
The authors wish to acknowledge the Irish Centre for High-End Computing (ICHEC) for the provision of computational facilities and support.
\ \
The authors acknowledge the EuroHPC Joint Undertaking for awarding this project access to the EuroHPC supercomputer Karolina, hosted by IT4Innovations through a EuroHPC Regular Access call (EHPC-REG-2023R03-103).
\ \
RSK and SCOG acknowledge financial support from the European Research Council via the ERC Synergy Grant ``ECOGAL'' (project ID 855130),  from the German Excellence Strategy via the Heidelberg Cluster of Excellence (EXC 2181 - 390900948) ``STRUCTURES'', and from the German Ministry for Economic Affairs and Climate Action in project ``MAINN'' (funding ID 50OO2206). 
\ \
RSK is grateful for computing resources provided by the Ministry of Science, Research and the Arts (MWK) of the State of Baden-W\"{u}rttemberg through bwHPC and the German Science Foundation (DFG) through grants INST 35/1134-1 FUGG and 35/1597-1 FUGG, and also for data storage at SDS@hd funded through grants INST 35/1314-1 FUGG and INST 35/1503-1 FUGG.
\ \
RSK also thanks the Harvard-Smithsonian Center for Astrophysics and the Radcliffe Institute for Advanced Studies for their hospitality during his sabbatical, and the 2024/25 Class of Radcliffe Fellows for highly interesting and stimulating discussions.

\end{acknowledgements}

\bibliographystyle{mnras}
\bibliography{references.bib}

\end{document}